\documentclass{llncs}

\newif\ifcomments
\commentstrue

\newif\ifanon
\anonfalse

\title{Optimized Point Addition Circuits for Elliptic Curve Discrete Logarithms}

\ifanon \else
\author{André Schrottenloher}
\institute{Univ Rennes, Inria, CNRS, IRISA, Rennes, France \\ \email{firstname.lastname@inria.fr}}
\fi

\usepackage{braket}
\usepackage{tikz}
\usepackage{algorithm}
\usepackage{algorithmicx}
\usepackage{algpseudocode}
\usepackage{mathtools}

\usepackage{multicol}
\usepackage{amsmath}
\usepackage{amssymb}

\usepackage{tikz}
\usetikzlibrary{decorations.pathreplacing,calc}

\spnewtheorem{assumption}{Assumption}{\itshape}{\rmfamily}

\usepackage{xifthen}

\newcommand{\bigO}[1]{\mathcal{O}
\ifthenelse{\isempty{#1}}{}{\! \left( #1 \right)}}
\newcommand{\bigOt}[1]{\widetilde{\mathcal{O}}
\ifthenelse{\isempty{#1}}{}{\! \left( #1 \right)}}

\usepackage[colorlinks]{hyperref}
\usepackage{booktabs}

\algblockdefx[QAA]{QAA}{EndQAA}%
[1]{\textbf{Run QAA over #1:}}%
{\textbf{EndQAA}}

\usepackage{tikz}
\usepackage{pgfplots}
\usetikzlibrary{calc,patterns,angles,quotes}
\usetikzlibrary{shapes.misc,positioning}
\tikzstyle{block}=[draw,minimum size=2em]
\tikzstyle{l}=[minimum size=1.5em]
\tikzstyle{xor}=[draw, circle, minimum size=0.8em,append after command={[shorten >=\pgflinewidth, shorten <=\pgflinewidth,] 
        (\tikzlastnode.north) edge (\tikzlastnode.south)
        (\tikzlastnode.east) edge (\tikzlastnode.west)
}]
\tikzstyle{sponge}=[rectangle, rounded corners=.25cm, minimum width=.5cm, minimum height=1.8cm, draw]
\tikzstyle{slash}=[append after command={(\tikzlastnode.south west) edge (\tikzlastnode.north east)}]
\tikzstyle{arrow}=[pin edge={to-,thin,black}]
\tikzstyle{point}=[minimum size=0.3em, below=-2pt,fill=black,circle]
\tikzstyle{trace}=[below=0.3em,pin edge={-, thin,black}, black]
\tikzstyle{plus}=[draw, thin, minimum size=0.8em,append after command={[shorten >=\pgflinewidth, shorten <=\pgflinewidth,] 
        (\tikzlastnode.north) edge (\tikzlastnode.south)
        (\tikzlastnode.east) edge (\tikzlastnode.west)
}]
\tikzstyle{rplus}=[draw, thin, minimum size=0.8em,append after command={[shorten >=\pgflinewidth, shorten <=\pgflinewidth,] 
        (\tikzlastnode.north) edge[color=red] (\tikzlastnode.south)
        (\tikzlastnode.east) edge[color=red] (\tikzlastnode.west)
}]
\tikzstyle{rpoint}=[minimum size=0.3em, below=-2pt,fill=red,circle]
\tikzstyle{rarrow}=[pin edge={to-,thin,red}]

\newcommand{\FF}{\mathbb{F}}
\renewcommand{\vec}[1]{\mathbf{#1}}

\begin{document}

\newcommand{\appendixautoref}[1]{Appendix~\ref{#1}}
\def\algorithmautorefname{Algorithm}
\def\problemautorefname{Problem}
\def\assumptionautorefname{Assumption}
\def\subsectionautorefname{Section}
\def\sectionautorefname{Section}

\maketitle

\begin{abstract}
Shor's algorithm represents the main threat of quantum computers to cryptography. In order to precisely understand its feasibility, many authors have worked towards reducing its costs, either at the logical level (assuming a fault-tolerant architecture), or at the physical level (taking into account the constraints of envisioned hardware). In particular, recent works by Chevignard et al. (CRYPTO 2024) and Gidney (arXiv 2025) used improved arithmetic to significantly reduce the qubit cost of factoring RSA public keys.

Even more recently, Babbush et al. (arXiv 2026) improved the cost of computing elliptic curve discrete logarithms, with a reduction of a factor 2 to 3 in gate count and qubit count compared to a previous work by Litinski (arXiv 2023). Their result relies on optimized point addition circuits on elliptic curves over prime fields. However they did not reveal their logical quantum circuits, relying instead on a zero-knowledge proof.

In this paper, we detail a quantum logical circuit architecture which gives similar results as Babbush et al., with a slightly higher number of qubits (around 1.5\% increase) and a slightly smaller Toffoli gate count (between 6.5\% and 10\% reduction) for the curve secp256k1. We also give gate counts for a generic variant of the circuit, which is valid for any prime field.
\end{abstract}


%================================

\section{Introduction}

Since Shor's breakthrough paper~\cite{DBLP:journals/siamcomp/Shor97}, it has been known that once a large-scale quantum computer becomes available, public-key cryptography based on the hardness of integer factoring (e.g., the RSA cryptosystem) and discrete logarithms (e.g., cryptography based on finite fields~\cite{NIST-SP800-56A-Rev3} and elliptic curve groups~\cite{NIST-SP800-56A-Rev3}) can be considered as broken, the underlying problems being solved in quantum polynomial time.

This impact motivates the early adoption of post-quantum cryptography, with a first batch of public-key encryption and signature schemes standardized by the NIST as a result of a large-scale public standardization process~\cite{nistcall}.

After Shor's paper, many authors have quantified the \emph{exact} gate and qubit counts of the algorithm, especially for its main cryptographic applications: factoring RSA integers, computing discrete logarithms in finite fields, and in elliptic curves of prime order. This paper focuses on the latter. Most of these studies~\cite{DBLP:journals/qic/ProosZ03,DBLP:conf/pqcrypto/HanerJNRS20,cryptoeprint:2026/280} have considered logical quantum circuits, that is, without any implementation constraints. As the technology has matured, several works have also examined the cost of physically implementing the algorithm under specific assumptions regarding the capabilities and scale of the available hardware. Both logical algorithms and physical mappings have advanced in parallel. As an example, in 2021, Gidney and Eker{\aa}~\cite{gidney2021factor} estimated that 20 million superconducting qubits could be necessary to factor RSA-2048 public keys (classically secure). A more recent paper by Gidney~\cite{gidney2025factor} reached one million qubits. Roughly half of this $20\times$ reduction (i.e., a factor 4) comes from logical circuits based on (and improving over)~\cite{DBLP:journals/iacr/ChevignardFS24}, while the rest comes from an improved physical implementation.

\paragraph{Babbush et al.'s Result.}
In a recent preprint, Babbush et al.~\cite{babbush2026securing} improved logical quantum circuits for computing elliptic curve discrete logarithms, and discussed the risks for cryptocurrencies using this non-post-quantum cryptography. In this paper, we only focus on their logical circuit results. 

Babbush et al.'s result is based on an improved (windowed) elliptic curve point addition circuit on the curve \texttt{secp256k1}. Their circuits are ``kickmick'' circuits, which allow measurement-based uncomputation and phases but are ultimately efficient to simulate classically. They did not publish any detail on their circuit architecture, relying instead on a zero-knowledge proof to prove that they obtained circuits with the claimed gate and qubit counts, which performed point addition with a reasonable success probability. Succeeding on random inputs is enough to run Shor's algorithm.

The decision not to publish the circuit is justified in~\cite{babbush2026securing} by concerns on the feasibility of such quantum attacks in practice on the medium term. This paper follows instead the practice of previous works~\cite{litinski2023compute,gouzien2023performance,gidney2025factor}, which intend to make the result fully reproducible. We believe that a full discussion of the arguments in~\cite{babbush2026securing} would be out of scope of the current paper.

\paragraph{Our Result.}
We obtain two circuits for point addition on \texttt{secp256k1}, reaching similar results as Babbush et al.~\cite{babbush2026securing}, with a slight improvement in Toffoli gate count. The design principles of these circuits are described in this paper. We programmed and tested them using Qarton\footnote{\url{https://gitlab.inria.fr/capsule/qarton}}. In particular, the circuits are not exact: they will fail with some probability. Testing on random inputs shows that this probability of failure is comparable to the one of~\cite{babbush2026securing}.

While the gate count varies depending on implementation details, the number of qubits is around $4.355n + \bigO{\sqrt{n}}$ in the version that we selected, where $n$ is the bit-size of the prime. It could be reduced up to $4.12n + \bigO{\sqrt{n}}$ using a different implementation, which would likely increase the Toffoli gate count. Another recent work~\cite{cryptoeprint:2026/280} achieved a space of $3.12n + \bigO{\sqrt{n}}$, but the technique employed is very different and yields a sharp increase in gate count for a relatively small gain in memory on relevant instances.

A comparison between the qubit and gate counts of Babbush et al.~\cite{babbush2026securing} and ours is given in~\autoref{tab:comparison}.

%* A space-optimized circuit with 1193 qubits and 2385517 = 2^21.19 Toffoli gates
%* A gate-optimized circuit with 1446 qubits and 1865521 = 2^20.83 Toffoli / AND gates

\begin{table}[htb]
\caption{Costs of our circuits and comparison with~\cite{babbush2026securing}. The ``Toffolis'' column counts together CCX, CCZ as well as And gates. All circuits have a failure probability at most $2^{-13.3}$ (succeeding on 10~000 randomly chosen inputs). The number of qubits and gates given here do not account for the small window of size $w = 16$: it costs 16 qubits more and $3 \times 2^w$ Toffoli gates to run the actual ``windowed point addition'' circuit.}\label{tab:comparison}
\centering
\begin{tabular}{lcccc}
\toprule
Type & Qubits & Toffolis & Total gates & Reference \\
\midrule
Space-optimized for \texttt{secp256k1} & 1175 & $2^{21.36}$ & $\simeq 2^{24}$ & \cite{babbush2026securing} \\
Gate-optimized for \texttt{secp256k1} &1425 & $2^{21.00}$ & $\simeq 2^{24}$ & \cite{babbush2026securing} \\
\midrule
Space-optimized for \texttt{secp256k1} & 1192 & $2^{21.19}$ & $\simeq 2^{24}$ & Ours \\
Gate-optimized for \texttt{secp256k1} & 1446 & $2^{20.83}$ & $\simeq 2^{24}$ & Ours \\
\midrule
Space-optimized for any prime & 1192 & $2^{21.78}$ & $\simeq 2^{24.5}$ & Ours \\
Gate-optimized for any prime & 1446 &  $2^{21.42}$ & $\simeq 2^{24.5}$ & Ours \\
\bottomrule
\end{tabular}
\end{table}

Following the same reasoning as~\cite{babbush2026securing}, the Toffoli count of the entire Shor's algorithm on \texttt{secp256k1} is roughly:
\begin{equation}
28 \times (3 \times 2^{16} + Q_A)
\end{equation}
where $Q_A$ is the Toffoli count of a single windowed addition circuit as given in~\autoref{tab:comparison}, while the qubit count is the one of~\autoref{tab:comparison} plus 16 (roughly). The algorithm needs to run only once to retrieve the discrete logarithm (contrary to trade-offs for RSA like in~\cite{gidney2025factor}, which require multiple runs). The resulting numbers are summarized in~\autoref{tab:shor}. The best previous result was from Litinski~\cite{litinski2023compute} using around 200 million Toffoli gates (i.e., $2^{27.57}$) to compute a single discrete logarithm, with about twice as much qubits.

\begin{table}[htb]
\caption{Cost of running Shor's algorithm entirely and comparison with~\cite{babbush2026securing}. The ``Toffolis'' column counts together CCX, CCZ as well as And gates.}\label{tab:shor}
\centering
\begin{tabular}{lcccc}
\toprule
Type & Qubits & Toffolis & Reference \\
\midrule
Space-optimized for \texttt{secp256k1} & 1191 & $ 28 \times (2^{21.36} + 3 \cdot 2^{16}) = 2^{26.27}$ & \cite{babbush2026securing} \\ 
Gate-optimized for \texttt{secp256k1} & 1441 & $ 28 \times (2^{21.00} + 3 \cdot 2^{16}) = 2^{25.94}$ & \cite{babbush2026securing} \\
\midrule
Space-optimized for \texttt{secp256k1} & 1208 & $ 28 \times (2^{21.19} + 3 \cdot 2^{16}) = 2^{26.11}$ & Ours \\
Gate-optimized for \texttt{secp256k1} & 1462 & $ 28 \times (2^{20.83} + 3 \cdot 2^{16}) = 2^{25.78}$ & Ours \\
\midrule
Space-optimized for any prime & 1208 & $ 28 \times (2^{21.78} + 3 \cdot 2^{16}) = 2^{26.66}$ & Ours \\
Gate-optimized for any prime & 1462 &  $ 28 \times (2^{21.42} + 3 \cdot 2^{16}) = 2^{26.33}$ & Ours \\
\bottomrule
\end{tabular}
\end{table}

\subsection*{Ideas of the Circuit and Organization of the Paper}

We detail our circuits in a top-down approach, starting from the high-level implementation of Shor's algorithm (which is the same as in~\cite{babbush2026securing} and previous works), down to the implementation of arithmetic circuits.

\paragraph{From Shor's Algorithm to Point Addition (\autoref{sec:high-level}).}
The fact that Babbush et al.~\cite{babbush2026securing} used a standard implementation of Shor's algorithm, and not a ``compressed'' version like~\cite{cryptoeprint:2026/280}, is explicit in their work. In this implementation, the main costly component in the algorithm is point multiplication:
$$ \ket{u,v} \ket{0} \mapsto \ket{u,v} \ket{[u] P + [v] Q} \enspace, $$
where $P,Q$ are points on the elliptic curve given by the discrete logarithm instance ($Q$ is the input and $P$ is the basis of the discrete logarithm). Although the registers $u,v$ take $n$ bits each roughly, where $n$ is the bit-size of the group of points, they do not count towards the total qubit count thanks to the \emph{semi-classical Fourier transform} with qubit recycling, and we can focus instead on the points.

Point multiplication is then reduced to \emph{windowed point addition}, where one uses a $w$-bit integer input $i$ to select a point from a built-in \emph{window} $(P_0, \ldots, P_{2^w-1})$, and add the constant point to the variable input:
$$ \ket{i} \ket{R} \mapsto \ket{i} \ket{R + P_i} \enspace. $$
Finally, the point addition by a constant is done by an algorithm of~\cite{gouzien2023performance}. Points are represented using \emph{affine} coordinates, i.e., a pair of integers modulo $q$. The addition by $P_i$ is reduced to a sequence of arithmetic operations modulo $q$ and three \emph{lookups}, which are the circuits loading the coordinates of $P_i$ from the input $i$. All this high-level structure is detailed in~\autoref{sec:high-level} and is explicit in Babbush et al.'s paper~\cite{babbush2026securing}.

\paragraph{From In-Place Multiplication to the Extended Euclidean Algorithm (\autoref{sec:modmul}).}
When doing point addition like this, one observes that the most costly step is to \emph{multiply in-place} two integers modulo $q$: $\ket{x,y} \mapsto \ket{x, xy}$.
This operation happens twice in the point addition circuit. Previous works like~\cite{DBLP:journals/qic/ProosZ03,DBLP:conf/pqcrypto/HanerJNRS20,gouzien2023performance,litinski2023compute} reduced it to \emph{modular inversion} and out-of-place multiplication ($\ket{x,y,z} \mapsto \ket{x,y,z + xy}$), and implemented modular inversion using reversible variants of binary Extended Euclidean Algorithm (EEA). While inversion and multiplication on $n$-bit modular integers have both a gate count in $\bigO{n^2}$ (as does the entire point addition circuit), the constant is noticeably bigger for the inversion part. 

In~\autoref{sec:modmul}, we detail the implementation of this step. The main idea, taken from~\cite{khattar2025verifiable}, is to separate the EEA into two sub-circuits. 
The first circuit is an Euclidean algorithm which operates on a pair of integers $(u,v)$ (which should start at $(q, x)$ if we are inverting $x$ modulo $q$), performs the operations of the Euclidean algorithm, and outputs a bit-string describing the operations that it has performed, ensuring reversibility. This sequence of operations is named a \emph{dialog} in~\cite{khattar2025verifiable}, which used this for polynomials, because it can contain a division by 2, an inversion (swapping the integers), or an addition. The second circuit is a \emph{Bézout reconstruction} algorithm which takes as input the sequence of operations from the Euclidean algorithm, as well as a pair $(r,s)$ (which should start at $(1,0$) if we are inverting $x$ modulo $q$), and applies the appropriate operations on $(r,s)$. Essentially, this two-step process is splitting the EEA between the input part and the Bézout coefficients. This has many advantages which are detailed in~\autoref{sec:modmul}: not only is the space reduced, but this point of view allows to directly implement the in-place multiplication, saving several modular multiplications.

The total space complexity of the point addition ends up being dominated by the Bézout reconstruction step. As we observe in~\autoref{sec:modmul}, in the most space-optimized implementation, the sequence of steps of the EEA can be stored on $2.12n + \bigO{\sqrt{n}}$ qubits. During the Bézout reconstruction step, one needs to store in addition a pair of integers modulo $q$, of $n$ bits each. Therefore the circuit uses at least $4.12n + \bigO{\sqrt{n}}$ qubits. In our implementation, we use slightly more space than this, leading to our $4.36n + \bigO{\sqrt{n}}$ asymptotic. One should note that the constant in the $\bigO{}$ controls the success probability of the algorithm on random inputs, and we optimize it following experimental results.

\paragraph{Arithmetic Circuits.}
At this point, the cost of the point addition circuit will depend on trade-offs in the circuits for unsigned arithmetic (in the Euclidean algorithm) and modular arithmetic (in the Bézout reconstruction and elsewhere). For unsigned arithmetic, we simply switch between the CDKM adder~\cite{cuccaro2004new}, the Gidney adder~\cite{DBLP:journals/quantum/Gidney18} and a hybrid between the two, depending on the number of ancilla qubits available. We also rely on Gidney's low-space classical-quantum adder~\cite{gidney2025classical}. For modular arithmetic, we introduce two levels of optimization which, to the best of our knowledge, have not been described before, and could have applications in other quantum cryptanalysis circuits relying on modular arithmetic. Contrary to most recent works~\cite{DBLP:conf/pqcrypto/HanerJNRS20,gouzien2023performance,litinski2023compute}, we do not rely on the Montgomery representation of modular integers, instead optimizing the circuits as much as possible under a standard representation of integers. The components that we optimize are modular doubling and controlled modular addition, taken from~\cite{roetteler2017quantum}. 

%In~\cite{roetteler2017quantum}, modular doubling costs $2n + o(n)$ Toffoli gates on $n$-bit inputs, and modular addition costs 

We give a first batch of optimizations working for any modulus $q$, where comparisons are simplified by comparing the most significant bits only: the amount of MSBs determines the success probability of the circuit. Then, we observe that the prime in \texttt{secp256k1} is a \emph{pseudo-Mersenne} prime, of the form $2^u - f$ where $f \ll 2^u$. This allows to further simplify the modular reductions which appear in these circuits. The difference between these two levels of optimization explains the difference between \texttt{secp256k1} and ``any prime'' reported in~\autoref{tab:comparison}.

\paragraph{Implementation.}
We provide a complete implementation of the circuits described in this paper using the python library Qarton. Our code is available at:
\begin{center}
\url{gitlab.inria.fr/capsule/qarton-projects/ec-point-addition}
\end{center}
Using this code, one can for example build the point addition circuit, simulate it on random inputs and check that the error probability is small. We note that for efficiency reasons, the simulation does not descend at the gate level. Instead it replaces \emph{exact} arithmetic circuits (addition, subtraction, comparison\dots) by their classical functions.

\section{High-Level Structure}\label{sec:high-level}

This section details the reduction from Shor's algorithm for discrete logarithms to the \emph{windowed point addition} circuit on elliptic curves, and gives the high-level structure of the circuit.

Let $E(\FF_q)$ be an elliptic curve defined over a prime field, where $q$ is a prime number of bit-size $n$ ($q < 2^n$). The curve is defined by an equation $y^2 = x^3 + ax + b$. In particular, the \texttt{secp256k1} curve uses:
\begin{equation*}
q = 2^{256} - 4294968273, a = 0, b = 7 \enspace.
\end{equation*}

A point on the curve in \emph{affine coordinates} is either of the form $(x,y)$, or the \emph{point at infinity} $\mathcal{O}$. From an implementation perspective, these representations can be unified by a triple $(x,y,b) \in \FF_q \times \FF_q \times \{0,1\}$ where $b = 0$ and $x = y = 0$ if the point is $\mathcal{O}$ and $b = 1$ otherwise. Points on the curve form an additive group where $\mathcal{O}$ is the neutral element.

\paragraph{Quantum Circuits.}
We describe quantum algorithms in the quantum circuit model. We refer to~\cite{nielsen2002quantum} for an overview, and assume familiarity with the notions of gates, qubits, and the ket notation $\ket{\cdot}$ of quantum states.

We focus on the optimization of \emph{logical} quantum circuits, without taking into account the error correction layer of fault-tolerant quantum architectures. At the logical layer, we are interested in minimizing the number of logical qubits, and the gate count, especially the non-Clifford gates, due to the higher difficulty of implementing them in fault-tolerant quantum computing. In previous works, Toffoli gates or T gates have been used as the main metric. In this paper, we use circuits which contain Toffoli gates or AND gates, which are decomposed differently in the Clifford+T gate set. Both are counted together as a single ``Toffoli gate'' metric. These are the only non-Clifford gates in the circuits. Other gates include not (X), controlled not (CX), Hadamard, phase flips (CZ).

All arithmetic circuits described in this paper can be efficiently simulated, despite using specific quantum effects like measurement-based uncomputation. Babbush et al.~\cite{babbush2026securing} notice this for their circuits, and describe a class of ``kickmix'' circuits which operate using classical gates, phases and measurements. More generally, the property we require (which is satisfied throughout this paper) is that simulating the circuit, starting from any valid input (a single basis state), produces a quantum state with a sparse support, having always at most a constant number of basis states. With measurement-based uncomputation, like in~\cite{gidney2025classical}, one typically applies Hadamard gates followed immediately by measurements, so the quantum state remains supported by two basis states only.

\paragraph{Shor's Algorithm.}
Let $P$ be a base point and $Q$ be the point of which we compute the discrete logarithm, i.e., $Q = [k] P$ for some unknown $k$. We use a version of Shor's algorithm that can be adapted with the semiclassical Fourier transform~\cite{griffiths1996semiclassical}:
\begin{enumerate}
\item Create a uniform superposition over two registers of size $n$: $\sum_{0 \leq u,v < 2^n} \ket{u} \ket{v}$
\item Apply a \emph{point multiplication unitary}: $\sum_{0 \leq u,v < 2^n} \ket{u} \ket{v} \ket{[u] P + [v]Q}$
\item Apply a $2^n$-dimensional Fourier transform over the two input registers
\item Measure
\end{enumerate}
One can then retrieve $k$ from a single measurement result with large success probability (more than 1/2). The \emph{semiclassical Fourier transform} interleaves the operations of the Fourier transform and the point multiplication using a single recycled control qubit. As a consequence, both qubit and gate count can be approximated by those of the point multiplication:
\begin{equation}
\ket{u,v} \ket{0} \mapsto \ket{u,v} \ket{[u] P + [v]Q} \enspace.
\end{equation}

%The registers for $u,v$ can be taken of size $\log_2 q$ each, i.e., 256 bits for \texttt{secp256k1}~\cite{babbush2026securing}. In addition, we use \emph{controlled windowed multiplications} to optimize the cost. 

Let $w$ be a \emph{window} size, which can be selected depending on the cost of point addition. The value is $w = 16$ in~\cite{babbush2026securing}, and since we have similar costs, we will use the same later on. We rewrite $u$ and $v$ as:
\begin{equation}
\begin{cases}
u := u_0 + 2^w u_1 + 2^{2w} u_2 + \ldots + 2^{(\ell-1) w} u_{\ell-1} \\
v := v_0 + 2^w v_1 + 2^{2w} v_2 + \ldots + 2^{(\ell-1) w} v_{\ell-1}
\end{cases}
\end{equation}
Then the point multiplication is rewritten as:
\begin{equation}
[u] P + [v]Q = \sum_{i=0}^{\ell-1} [u_i] ([2^{iw}]P) + \sum_{i=0}^{\ell-1} [v_i] ([2^{iw}]Q) \enspace.
\end{equation}

We can therefore reduce the point multiplication circuit to the following \emph{windowed point addition}. Let $(P_0, \ldots, P_{2^w-1})$ be a \emph{window} of points, then the point addition circuit does:
\begin{equation}
\ket{i} \ket{R} \mapsto \ket{i} \ket{R + P_i} \enspace.
\end{equation}

Starting from $\mathcal{O}$, we just have to compute $2\ell = (2n/w)$ times a windowed point addition circuit, where the windows are precomputed multiples of $P$ and $Q$. If $i = 0$ the multiple is $\mathcal{O}$. The semiclassical Fourier transform works similarly for windowed additions, with less intermediate measurements since we only have to interleave the operations between each window.

\paragraph{Point Addition.}
The point addition circuit uses the strategy of~\cite{gouzien2023performance}. We can use formulas for addition of elliptic curve points in affine coordinates, neglecting the case of doublings (which would happen only with negligible probability) and the point at infinity. The only special case is for $i = 0$, where the output register should be unchanged since we are adding $\mathcal{O}$. 
%In fact, we can also include at almost no overhead the case where the input point $R$ is $\mathcal{O}$. In that case, we just write $P_i$ into the output register, as $\mathcal{O}$ is neutral for addition.

During the computation of the formulas, we need to use the coordinates of $P_i$ three separate times: first the entire point, then $3 x_{P_i}$, then the entire point again. Each time, we load these coordinates using a \emph{table lookup} circuit, and unload them immediately after use. A table lookup of $2^w$ values costs $2^w$ Toffoli gates~\cite{babbush2018encoding}, and the unlookup is almost free thanks to measurement-based uncomputation. In addition, the Toffoli gate count of lookups being still an order of magnitude below the cost of the other operations, we could also just compute the inverse of the lookup circuit.

%        # Formulas to follow:
%        # lambda = (y1-y2) / (x1 - x2) = (y1 + y3)/(x1-x3)
%        # x3 = lambda^2 - x1 - x2
%        # y3 = lambda(x1-x3) - y1

As can be seen in~\autoref{algo:point-addition}, the circuit contains 3 lookups, some arithmetic operations in $\bigO{n}$ gates, a modular squaring, and two \emph{in-place multiplication} circuits which we will detail next.

\begin{algorithm}[tb]
\caption{Reversible point addition circuit (from~\cite{gouzien2023performance}, slightly modified to allow addition of $\mathcal{O}$).}\label{algo:point-addition}
\begin{algorithmic}[1]
\Statex \textbf{Parameter:} window of points $P_0, \ldots, P_{2^w-1}$
\Statex \textbf{Input:} Registers $x_2, y_2$ (coordinates of $R$), register $i$
\Statex \textbf{Output:} Registers are modified in-place and contain $R + P_i$

\State $c \leftarrow (i \neq 0)$
\State $x_1, y_1 \leftarrow \text{Lookup}(P_i)$ \Comment{First lookup}
\Statex \Comment{When $i = 0$, we have $x_1,y_1 = 0$ per our definition}
\State $x_2 \leftarrow x_2- x_1$
\State $y_2 \leftarrow y_2- y_1$
\State $\text{Unlookup}(P_i)$ \Comment{Removes $x_1,y_1$}
\State $x_2, y_2 \leftarrow x_2, y_2 \cdot x_2^{-1}$  \Comment{Inverse of an in-place multiplication}
\State $x \leftarrow \text{Lookup}(3x_{P_i})$ \Comment{Second lookup}
\State $x_2 \leftarrow x_2 + x$
\State $\text{Unlookup}(3x_{P_i})$ \Comment{Removes $x$}
\State If $c$ then $x_2 \leftarrow x_2 - y_2^2$ \label{step:square} \Comment{Controlled modular squaring}
\State $x_2, y_2 \leftarrow x_2, y_2 \cdot x_2$ \Comment{In-place multiplication}
\State If $c$ then $x_2 \leftarrow -x_2$ \Comment{Controlled modular negation}
\State $x_1, y_1 \leftarrow \text{Lookup}(P_i)$ \Comment{Third lookup}
\State $y_2 \leftarrow y_2 - y_1$
\State $x_2 \leftarrow x_2 + x_1$
\State $\text{Unlookup}(P_i)$ \Comment{Removes $x_1,y_1$}
\State Uncompute $c$
\end{algorithmic}
\end{algorithm}

\paragraph{Number of Point Additions.}
The number of point additions can slightly be optimized~\cite{babbush2026securing}, following~\cite{litinski2023compute}. Indeed, the first addition can be replaced by a lookup of the result, and the last three ones can be removed at the cost of more classical post-processing~\cite{DBLP:journals/corr/abs-1905-09084}. In the \texttt{secp256k1} case, this means that only 28 point additions are necessary.

\section{Modular In-Place Multiplication}\label{sec:modmul}

The modular in-place multiplication:
\begin{equation}
\ket{x} \ket{y} \mapsto \ket{x} \ket{y x}
\end{equation}
is the core component of the circuit. Notice that the inverse circuit will multiply by $x^{-1}$ instead. So, using the same circuit, we can implement both in-place multiplication steps found in~\autoref{algo:point-addition}. The remaining multiplication, a modular squaring at Step~\ref{step:square}, is out-of-place, and therefore easier to implement.

We explain how to build this circuit from basic modular and non-modular arithmetic circuits. The idea is taken from~\cite{khattar2025verifiable}, which optimizes extended Euclidean algorithms (EEA) using \emph{dialog} representations. As we do not use the same representation, we will not use the term ``dialog'' but the idea remains the same.

\subsection{Euclidean Algorithm}

We start from a simple, but efficient, implementation of the Euclidean algorithm. The key of this implementation is that, contrary to the usual Extended Euclidean algorithm, it does not construct the Bézout coefficients $(r,s)$. Instead, it just remembers the choices that were made at each loop iterate, packing them into a \emph{garbage bit-vector}. Then, we can \emph{later} reconstruct $(r,s)$, gaining space compared to a direct implementation. This is the first key idea from~\cite{khattar2025verifiable}.

\begin{algorithm}[htb]
\caption{Euclidean Algorithm}\label{algo:ea}
\begin{algorithmic}[1]
\Statex \textbf{Parameters:} number of iterations
\Statex \textbf{Input:} Integers $u,v$, $u$ odd
\Statex \textbf{Output:} Garbage bit-vector $\vec{g}$

\For{A fixed number of iterations}
\State $b_0 := v \bmod{2}$
\State $b_1 := u > v$
\State Store $b_0 \& b_1$ and $b_0$ in $\vec{g}$
\If{$b_0 \& b_1$}
\State $u,v = v,u$
\EndIf
\If{$b_0$}
\State $v -= u$
\EndIf
\State $v = v/2$
\EndFor
\Comment{At the end, $v = 0$ and $u = 1$}
\end{algorithmic}
\end{algorithm}

\newcommand{\citer}{c_{\mathrm{iter}}}
\newcommand{\cpad}{c_{\mathrm{pad}}}

We review the important properties of~\autoref{algo:ea}. 

First, its number of iterations. Let $n$ be the starting bit-size of $u$ and $v$. Intuitively, each iteration will at least remove one bit of the registers at Step 11, and with probability 1/2 depending on the value of $b_0$, it will remove another bit at Step 9. This means that on average, $(uv)$ is multiplied by $\frac{1}{2} \times \frac{1}{2} + \frac{1}{2} \times \frac{1}{4} = \frac{3}{8}$. So, the average number of iterations should be $\ell$ such that: $2n - \ell \log_2(8/3) = 0 \implies \ell = \frac{2n}{3 - \log_2 3} \simeq 1.413n$.

Experiments show that this is indeed the case. In addition, the number of iterations follows a normal distribution with standard deviation $\simeq 0.6 \sqrt{n}$. In order to ensure a large success probability on random inputs, we will therefore choose an appropriate constant $\citer{} \simeq 2.4$ (four standard deviations) and use a fixed number of $1.413n + \citer{} \sqrt{n}$ iterations. 

Second, the number of qubits. At the beginning of the algorithm, $u$ and $v$ are stored on $2n$ qubits. Each iteration produces not two, but $1.5$ bits of garbage on average. Indeed, we only need to store $b_0 \& b_1$ if $b_0 = 1$. So the total number of output bits that can be expected is $1.5 \times (1.413 + \bigO{\sqrt{n}}) = 2.12n + \bigO{\sqrt{n}}$, and this is the minimal number of qubits with which we can run the algorithm. This situation is very similar to the Jacobi symbol algorithm in~\cite{cryptoeprint:2026/280}: the amount of reduction of $(u,v)$ and the number of garbage bits produced per iterate are the same.

In order to optimize the space, we use \emph{register sharing}. There exists a \emph{padding} amount $\cpad \sqrt{n}$, with $\cpad{} \simeq 2.3$, such that, with high probability, at each iteration $j$, the current values of $u$ and $v$ each fit on: $ n - 0.5 \log_2(8/3) i + \cpad \sqrt{n}$ bits. This is also experimentally observed. Therefore:
\begin{itemize}
\item Operations on $u$ and $v$ are performed on these sub-registers only;
\item We use the cleared qubits to store the garbage bits, ensuring a total space in $2.12n + \bigO{\sqrt{n}}$.
\end{itemize}
The choice of $\cpad{}$ and $\citer{}$ control the success probability of the algorithm when called on random inputs. According to our experiments, $\cpad{} = 2.3$ and $\citer{} = 2.4$ ensure a success probability large enough so that the point addition circuit succeeds on 10~000 random inputs.

\paragraph{Garbage Encoding.}
Compared to~\cite{cryptoeprint:2026/280}, we modify the storage of garbage bits. Previously we would have used a register in which $b_0$ and $b_0 \& b_1$ are queued using controlled-shifts. The issue is that the shifts are costly, and also, this makes the total size of the garbage vary, so we would need another $\bigO{\sqrt{n}}$ bits of space to absorb this variation. We use a simpler method here.

\begin{figure}[htbp]
\centering
\includegraphics[scale=0.8]{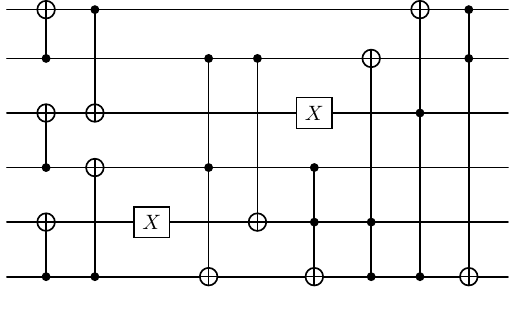}
\caption{``Compression'' circuit mapping 3 pairs $(b_0, b_0 \& b_1)$ into 5 bits. On valid inputs, the last bit is always 0 and can be reused.}\label{fig:compressor}
\end{figure}

Consider three successive iterations producing three pairs of bits. Since these bits are $(b_0, b_0 \& b_1)$ for some $b_0,b_1$, they can only be $(0,0)$, $(1,0)$ or $(1,1)$. There are $3^3 = 27 < 2^5$ possible values in input, so we can \emph{compress} this input value into 5 bits, releasing one bit. We define an ad-hoc quantum circuit which does this in-place with 5 Toffoli gates, shown in~\autoref{fig:compressor}.

With this, our Euclidean algorithm circuit uses exactly $(1.413n + \citer{} \sqrt{n})/3 \times 5 \simeq 2.355n + \bigO{\sqrt{n}}$ garbage bits. While this is asymptotically more, the compression step is much more efficient than the previous encoding (accounting at the end for a negligible cost), and the garbage register has now a fixed size.

\subsection{Using the Garbage Bit-Vector}

As we said above, in the EEA, one also maintains registers $(r,s)$ for the Bézout coefficints which are updated depending on $b_0$ and $b_0 \& b_1$. If we start the Euclidean algorithm with inputs $(u, v) := (q, x)$, and then run the updates on the Bézout coefficients $(r,s) := (1, 0)$, at the end of the algorithm we have $(r,s) = (0, x^{-1} \bmod{q})$.

The second key idea of~\cite{khattar2025verifiable} is that these updates are all \emph{linear} operations controlled by the bits $b_0$ and $b_0 \& b_1$. Therefore, if we start from $(r,s) := (y, 0)$ for any $y$ modulo $q$, we will end up with $(r,s) = (0, y x^{-1} \bmod{q})$: we have performed both the inversion \emph{and} an in-place multiplication without additional cost.

\begin{algorithm}[htb]
\caption{Bézout reconstruction algorithm.}\label{algo:dialog-reconstruction}
\begin{algorithmic}[1]
\Statex \textbf{Parameters:} prime $q$, number of iterations
\Statex \textbf{Input:} Integers $r, s$, garbage bit-vector $\vec{g}$
%\Statex \textbf{Output:} Integers $u,v$

%\State $u,v \leftarrow x,y$
\For{$i$ in reversed(range(iterations))}
\State Obtain $b_0$ and $b_0 \& b_1$ for iteration $i$ from $\vec{g}$
\State $s \leftarrow 2s \bmod{q}$
\If{$b_0$}
\State $s \leftarrow s + r \bmod{q}$
\EndIf
\If{$b_0 \& b_1$}
\State $r, s \leftarrow s, r$
\EndIf
\EndFor
\end{algorithmic}
\end{algorithm}

Our implementation follows~\autoref{algo:dialog-reconstruction}. We read the bits in the reversed order in which they were produced. The consequence is that, on input $(y,0)$, we obtain $(0, y x \bmod{q})$ and not $yx^{-1}$. Notice that in this reconstruction algorithm, we need modular arithmetic operations, contrary to the original EEA.

\begin{remark}
Running side by side~\autoref{algo:ea} and~\autoref{algo:dialog-reconstruction} essentially gives a reasonably efficient variant of binary GCD algorithm as used in previous implementations of elliptic curve point addition~\cite{litinski2023compute,gouzien2023performance}. However these implementations needed to run the algorithm once forwards, obtain a modular inverse, and run the algorithm backwards to erase the garbage. The out-of-band reconstruction method from~\cite{khattar2025verifiable} is also more efficient because only~\autoref{algo:ea} is run twice.
\end{remark}

We can now use~\autoref{algo:ea} and~\autoref{algo:dialog-reconstruction} to implement in-place modular multiplication. On input $(x,y)$, we have three steps:
\begin{itemize}
\item We convert $x$ into a garbage bit-vector $\vec{g}$ representing the EEA operations;
\item We use $\vec{g}$ to map $y$ into $xy \bmod{q}$;
\item We convert $\vec{g}$ back to $x$. 
\end{itemize}
The algorithm is~\autoref{algo:ipmul}.

\begin{algorithm}[htb]
\caption{In-place modular multiplication.}\label{algo:ipmul}
\begin{algorithmic}[1]
\Statex \textbf{Input:} Registers $x,y$
\State \textbf{Output:} Registers contain $x, yx \bmod{q}$

\State Convert $x$ to bit-vector $\vec{g}$ using~\autoref{algo:ea}
\Statex \Comment{Registers are now $\vec{g}, y$}
\State Use the Bézout reconstruction algorithm~\autoref{algo:dialog-reconstruction} on $(y,0)$
\Statex \Comment{Registers are now $\vec{g}, yx \bmod{q}$}
\State Convert back $\vec{g}$ to $x$
\Statex \Comment{Registers are now $x, yx \bmod{q}$}
\end{algorithmic}
\end{algorithm}

%Using space-efficient components for modular and non-modular arithmetic, we observe that the space complexity is dominated by the reconstruction step in~\autoref{algo:ipmul}: at this step, we need to store $2n$ bits for the two input registers, as well as the bit-vector representation of size $2.355n + \bigO{\sqrt{n}}$. This gives a total space complexity of $4.355 n + \bigO{\sqrt{n}}$ for in-place modular multiplication, transmitting to the point addition circuit.

We can make a few observations:
\begin{itemize}
\item The space complexity of the entire point addition circuit depends on~\autoref{algo:ipmul} only. All other steps can be implemented using space-efficient components, or sometimes using some trade-offs. For example, ancillas are available during the modular squaring.
\item If space-efficient components are used for arithmetic operations, then the dominating step in space complexity is the Bézout reconstruction step. At this point, we need to store $2n$ bits for the two input registers, as well as the bit-vector representation of size $2.355n + \bigO{\sqrt{n}}$. This gives a total space complexity of $4.355 n + \bigO{\sqrt{n}}$.
\item The GCD construction (\autoref{algo:ea}), both forwards and backwards, uses only $2.355n + \bigO{\sqrt{n}}$ qubits in total. With the additional register of $n$ bits stored at this point, this means that around $n$ qubits are available for trade-offs. In particular, we can use more gate-efficient arithmetic circuit like Gidney's adder~\cite{DBLP:journals/quantum/Gidney18}, which we cannot use during the reconstruction.
\item If we still use gate-efficient arithmetic during the reconstruction step, then we will need $n$ additional ancilla qubits, but we will save a few Toffoli gates. This is precisely the difference between the two variants of the Babbush et al.'s circuits~\cite{babbush2026securing} as well as ours.
\end{itemize}

\section{Approximate Arithmetic and Modular Arithmetic}\label{sec:approxarith}

At this point, we will obtain a circuit with roughly the same qubit count as~\cite{babbush2026securing} (a bit higher), but a larger gate count. We then optimize the arithmetic components to reduce the gate count as much as possible. There are three types of optimizations that we perform, which we detail below:
\begin{itemize}
\item Using appropriate arithmetic circuits. So far all the circuits used are exact and this has no influence on the success probability.
\item Using approximate arithmetic circuits for some steps. This has an influence on the success probability.
\item Using approximate modular arithmetic circuits taking advantage of the special form of the prime. This has an influence on the success probability, and makes our circuit dependent on the instance considered.
\end{itemize}

\paragraph{Trade-offs Between Implementations.}
When implementing arithmetic circuits with two inputs (adders, comparators, etc.), we commonly have the choice between two implementations:
\begin{itemize}
\item Implementations using (almost) no ancillas, based on the CDKM adder~\cite{cuccaro2004new} or variants. CDKM-based $n$-bit addition and comparison uses $2n$ Toffoli gates, $n$-bit controlled addition uses $3n$ Toffoli gates. 
\item Implementations using around $n$ ancillas using the Gidney adder~\cite{DBLP:journals/quantum/Gidney18}. This adder uses measurement-based uncomputation to uncompute AND gates efficiently. Addition uses $n$ Toffoli gates, controlled addition uses $2n$ Toffoli gates.
\end{itemize}
We use strategically components based on the latter when ancillas are available, such as in the modular squaring step, and components based on the former when this is not the case. Notably, during the execution of~\autoref{algo:ea} (or the inverse), we have access to roughly $n$ ancillas, so we can use the Gidney adder for the subtraction step. In fact, we do not have \emph{exactly} $n$ ancillas, because the circuit itself is using some of them as padding for the registers $u$ and $v$. Therefore, we used a hybrid construction as proposed in~\cite{DBLP:journals/quantum/Gidney18} for the subtraction step in~\autoref{algo:ea}, which uses exactly the amount of available ancillas.

\paragraph{Approximate Arithmetic Circuits.}
Shor's algorithm requires the point addition circuit to work with a large constant probability. This leaves us a lot of freedom to simplify the cost of arithmetic operations. In our circuits, we simplified all comparisons by comparing the most significant bits only (we typically used 40 to 50 bits to ensure a large probability of success). This simplification is used in the comparison step of~\autoref{algo:ea} and inside the modular arithmetic circuits.

\paragraph{Approximate Modular Arithmetic.}
The modular arithmetic circuits that we implement, both for the Bézout coefficient reconstruction (\autoref{algo:dialog-reconstruction}) and for the modular squaring step, are:
\begin{itemize}
\item (Controlled) Modular addition: $\ket{x,y} \mapsto \ket{x, y + x \bmod{q}}$
\item Modular doubling: $\ket{x} \mapsto \ket{2x \bmod{q}}$.
\end{itemize}

These circuits only need to work when their inputs are drawn uniformly at random, with a single exception. During the Bézout reconstruction (\autoref{algo:dialog-reconstruction}), for the first $\citer \sqrt{n}$ iterations, we have to handle the case where $x + y = q$. This should not appear if we pick the inputs at random, but it does appear in our case for the first empty iterations of the algorithm.

Our starting points for these two circuits are taken from~\cite{roetteler2017quantum}. We start by simplifying the comparisons as shown above, then we simplify them further using the special form of the prime $q$. In the following, we give the circuits as pseudocode which is similar to our python code. The commands used  in pseudocode are: \texttt{input} to allocate an input register (which is also output), of a given type, \texttt{anc} to allocate ancillas, basic gates, operations on registers such as \verb|shift_right|, \texttt{sub} (subtraction), \texttt{cadd} (controlled addition), \texttt{lt} (lower than). All these operations are in-place, and \texttt{lt} XORs the result of the comparison to its last input register. \texttt{clt} is the controlled version. We use ``+'' to denote concatenation of registers, seen as lists of qubits.

\paragraph{Doubling.}
On input $x$, the doubling circuit of~\cite{roetteler2017quantum} does a shift, a subtraction by $q$, and a controlled addition by $q$. This is given in~\autoref{algo:doubling}.

\begin{algorithm}[htb]
\caption{Exact modular doubling circuit from~\cite{roetteler2017quantum}.}\label{algo:doubling}
\begin{verbatim}
x_reg = input(ModIntType(q))
anc1, anc2 = anc(1), anc(1)
shift_right(x_reg + anc1, 1)
sub_uint(q, x_reg + anc1 + anc2)
cadd_uint(anc2, q, x_reg + anc1)
x(anc2)
cx(x_reg[0], anc2)
\end{verbatim}
\end{algorithm}

Our ``efficient'' modular doubling, which works for any prime, is given in~\autoref{algo:doubling-efficient}. The only non-exact step is the comparison, where we take the most significant bits of the current register and $q$, whereas the exact version of this circuit would have compared the entire numbers. We notice that the cost of modular doubling has been reduced to approximately one $n$-bit controlled constant addition.

\begin{algorithm}[htb]
\caption{Non-exact modular doubling circuit.}\label{algo:doubling-efficient}
\begin{verbatim}
x_reg = input(ModIntType(q))
anc1, anc2 = anc(1), anc(1)
shift_right(x_reg + anc1, 1)
lt_uint(x_reg[-padding:] + anc1, q_msbs, anc2) # non-exact step
self.x(anc2)
csub_uint(anc2, q, x_reg + anc1)
cx(x_reg[0], anc2)
\end{verbatim}
\end{algorithm}

When $q$ is a pseudo-Mersenne prime of the form $2^u-f$, where $f \ll 2^u$, we can further simplify this circuit. Indeed, we can further approximate the comparison by just checking whether the MSB is 1, and in that case, the controlled-subtraction by $q$ becomes an erasure of the MSB, followed by adding $f$ into the register $x$. Since $f$ is small, we only need to add it into the LSBs (since carries will only propagate a certain amount). The exact range determines the success probability of the algorithm. This is given in~\autoref{algo:doubling-special}.

\begin{algorithm}[htb]
\caption{Pseudo-Mersenne modular doubling circuit.}\label{algo:doubling-special}
\begin{verbatim}
x_reg = input(ModIntType(q))
anc1 = anc(1)
shift_right(x_reg + anc1, 1)
cadd_uint(anc1, f, x_reg[:lsbs])
cx(x_reg[0], anc1[0])
\end{verbatim}
\end{algorithm}

%That is, we shift, and if the MSB is 1, we add $f$ into the LSBs of $x$ (corresponding to a subtraction by $q$). Then we erase the MSB.

\paragraph{Addition.}
On input $x,y$, the controlled modular addition circuit of~\cite{roetteler2017quantum} does a controlled addition, followed by a subtraction of $q$, a controlled addition of $q$ and a comparison to erase an ancilla qubit. (Note that compared to the non-controlled circuit, the subtraction / addition of $q$ do not need to be modified).

\begin{algorithm}
\caption{Exact controlled modular addition circuit from~\cite{roetteler2017quantum}.}\label{algo:add}
\begin{verbatim}
control = input(BoolType())
x_reg = input(ModIntType(p))
y_reg = input(ModIntType(p))
anc_x = anc(1)
anc_y = anc(1)
cadd_uint(control, x_reg + anc_x, y_reg + anc_y)
sub_uint(p, y_reg + anc_y)
cadd_uint(anc_y, p, y_reg)
clt_uint(control, y_reg, x_reg, anc_y)
x(anc_y)
\end{verbatim}
\end{algorithm}

Our first simplification is given in~\autoref{algo:efficient-add}. We first perform a non-modular (controlled) addition, then check if the result overflows (but only on the MSBs). If it does, we subtract $q$ from the result. The ancilla bit checking the overflow can then be erased by the condition $y < x$, since the only case where this happens is precisely when there was an overflow. Both comparisons being performed only on the MSBs, the remaining dominating operations are a controlled addition and a controlled subtraction by a constant.

\begin{algorithm}
\caption{Non-exact controlled modular addition circuit.}\label{algo:efficient-add}
\begin{verbatim}
control = input(BoolType())
x_reg = input(ModIntType(p))
y_reg = input(ModIntType(p))
anc_x = anc(1)
anc_y = anc(1)
cadd_uint(control, x_reg + anc_x, y_reg + anc_y)
anc = anc(1)
lt_uint(y_reg[-msbs:] + anc_y, q_msbs, anc)
self.x(anc[0])
csub_uint(anc, q, y_reg + anc_y)
clt_uint(control,y_reg[-msbs:], x_reg[-msbs:], anc)
\end{verbatim}
\end{algorithm}

When $q$ is a pseudo-Mersenne prime of the form $2^u - f$, where $f \ll 2^u$, we can further simplify this circuit. Here we need two variants: one that handles the case $x + y = q$ (which would not happen randomly), and one that does not need to.

\begin{algorithm}
\caption{Pseudo-Mersenne controlled modular addition circuit, which does not handle $x + y = q$.}\label{algo:special-add}
\begin{verbatim}
control = input(BoolType())
x_reg = input(ModIntType(q))
y_reg = input(ModIntType(q))
anc_x = anc(1)
anc_y = anc(1)
cadd_uint(control, x_reg + anc_x, y_reg + anc_y)
cadd_uint(anc_y, f, y_reg[:lsbs])
clt_uint(control, y_reg[-msbs:], x_reg[-msbs:], anc_y)
\end{verbatim}
\end{algorithm}

The latter is given in~\autoref{algo:special-add}. We first perform the controlled addition, then if the register overflows, we add $f$ in the LSBs of the output register $y$. We erase the overflow bit by a new comparison.

The algorithm handling the case $x + y = q$ is given in~\autoref{algo:special-add-handleq}. It detects whether $x+y = q$ by checking if all MSBs are 1. If $x + y = q$, then after reduction, the register $y$ should contain 0. Thus the bit is erased by checking if all MSBs are 0.

\begin{algorithm}
\caption{Pseudo-Mersenne controlled modular addition circuit which handles $x + y = q$.}\label{algo:special-add-handleq}
\begin{verbatim}
control = input(BoolType())
x_reg = input(ModIntType(q))
y_reg = input(ModIntType(q))
anc_x = anc(1)
anc_y = anc(1)
cadd_uint(control, x_reg + anc_x,y_reg + anc_y)
anc = anc(1)
y_is_q = all(y_reg[-msbs:])
cxor(y_is_q, q, y_reg) # Controlled-XOR of constant (CNOT gates)
cx(anc_y, anc)
cadd_uint(anc, f, y_reg[:lsbs])
cx(anc, anc_y)
clt_uint(control, y_reg[-msbs:], x_reg[-msbs:], anc)
cx(y_is_q, anc)
y_is_q = all(not y_reg[-msbs:])
\end{verbatim}
\end{algorithm}

\paragraph{Difference of Cost.}
As can be seen in~\autoref{tab:comparison}, there is a significant difference between the case of a pseudo-Mersenne prime as in \texttt{secp256k1} and the generic case. This comes entirely from the different implementations given above. An important factor is that, for the operations which happen during the Bézout reconstruction step, we need to optimize the space, so when we need to add or subtract $q$, we use Gidney's constant adder~\cite{gidney2025classical} with dirty ancillas (which costs $3n$ Toffoli gates for $n$-bit integers). Reducing the constant addition steps from $q$ to $f$ is a significant gain of gate count.

%If $q$ does not have the shape $2^u -f$, we cannot use the above optimizations (only the simplification of comparators). The main difference is that we have to perform space-efficient constant additions by $q$ or $-q$. One can use Gidney's construction~\cite{gidney2025classical} for this (the version with dirty ancillas, which costs $3n$ Toffoli gates, is the most adapted).

%\section{Discussion}
%The entirety of~\autoref{sec:high-level} (high-level structure of the algorithm) follows from existing works and is explicitly named in~\cite{babbush2026securing}. Therefore we believe that their improvement comes entirely from the modular in-place multiplication operation.
%
%It is well-known that the costly step in this circuit is the modular inverse. The ``compiled'' extended Euclidean algorithm using dialogs has been discussed in great detail in~\cite{khattar2025verifiable} for the case of polynomials, and applying the same idea on integers is natural.
%
%Likewise, register sharing and iteration optimization (e.g., we reduce the size of the registers $u,v$ at each iteration) play a large role in reducing the qubit and gate count respectively. These are known techniques which have been used repeatedly. When $q$ is a general prime, modular arithmetic steps can be optimized using Gidney's constant adder~\cite{gidney2025classical} (it did not play a large role in our case because we use the special form of $q$).

\section{Discussion}

After constructing the entire circuit, it is interesting to look at the proportion of the CCX cost of all its sub-circuits (we count together CCX and And gates resulting from Gidney's adder construction~\cite{DBLP:journals/quantum/Gidney18}). Our estimations are given in~\autoref{tab:proportion}. We can see as noticed above that the difference in gate count between a pseudo-Mersenne prime and a generic prime can be attributed to the modular addition and doubling circuits, and more precisely to the constant additions therein. 

%It follows from this estimation that further prime-specific optimizations may have a significant impact (as all the occurrences of Gidney's constant adder~\cite{gidney2025classical} are related to the prime). We also believe that prime-specific optimizations of modular arithmetic could be very useful in other contexts, like proofs of quantum advantage~\cite{DBLP:conf/stoc/Kahanamoku-Meyer25}.

\begin{table}
\centering
\caption{Proportion of the CCX cost of different sub-circuit components.}\label{tab:proportion}
\begin{tabular}{lrr}
\toprule
& \begin{tabular}{c} Space-optimized \\ for \texttt{secp256k1}\end{tabular} & \begin{tabular}{c} Space-optimized \\ for any prime \end{tabular} \\
\midrule
Main components: \\
\midrule
Modular squaring & 9 \%  &  10 \% \\
In-place multiplier and its inverse & 90 \% & 90 \% \\
Bézout reconstruction & 54 \%  & 64 \% \\
GCD construction & 36 \%  & 24 \% \\
\midrule
Some sub-components: \\
\midrule
Controlled-swap & 19 \% & 12 \% \\
Hybrid adder & 22 \%  & 14 \%  \\
%Among arithmetic circuits: \\
Controlled modular adder (and inverse) & 39 \% &  45 \% \\
Modular double (and inverse) & 8 \% & 24 \% \\
Gidney constant adder (controlled and not) & 15 \% & 34 \% \\
\bottomrule
\end{tabular}
\end{table}

\ifanon
\else
\subsubsection*{Acknowledgments.}

The author thanks Pierre-Alain Fouque, Dahmun Goudarzi, Amaury Pouly, and Yixin Shen for motivating discussions.
% Tanuj Khattar, 
%Supported by Decrypt ANR, like everything.
This work has been supported by the French Agence Nationale de la Recherche through the France 2030 program under grant agreement No. ANR-22-PETQ-0008 PQ-TLS

\fi

%\appendix

\bibliographystyle{splncs04}
\bibliography{biblio}

% nothing here

\end{document}